\documentclass[superscriptaddress]{revtex4-2}
\usepackage{graphicx}
\usepackage[dvipsnames]{xcolor}
\usepackage{bm}
\usepackage{siunitx}
\usepackage{amsmath}
\usepackage{color}
\usepackage{xr}
\usepackage[colorlinks=true,allcolors=blue]{hyperref}
\usepackage[normalem]{ulem}

\renewcommand{\vec}[1]{\bm{#1}}
\newcommand{\uvec}[1]{\hat{\bm{#1}}}

\begin{document}

\author{A.~S.~Meeussen}
\affiliation{School of Engineering and Applied Sciences, Harvard University, Cambridge,	Massachusetts 02138, USA}
\author{G.~Bordiga}
\affiliation{School of Engineering and Applied Sciences, Harvard University, Cambridge,	Massachusetts 02138, USA}
\author{A.~X.~Chang}
\affiliation{School of Engineering and Applied Sciences, Harvard University, Cambridge,	Massachusetts 02138, USA}
\author{B.~Spoettling}
\affiliation{ETH Zuerich, Zuerich 8092, Switzerland}
\author{K.~P.~Becker}
\affiliation{School of Engineering and Applied Sciences, Harvard University, Cambridge,	Massachusetts 02138, USA}
\affiliation{Department of Mechanical Engineering, Massachusetts Institute of Technology, Cambridge, Massachusetts 02139, USA}
\author{L.~Mahadevan}
\affiliation{School of Engineering and Applied Sciences, Harvard University, Cambridge,	Massachusetts 02138, USA}
\author{K.~Bertoldi*}
\affiliation{School of Engineering and Applied Sciences, Harvard University, Cambridge,	Massachusetts 02138, USA}

\title{Textile hinges enable extreme properties of mechanical metamaterials}

\begin{abstract}
Mechanical metamaterials---structures with unusual properties that emerge from their internal architecture---that are designed to undergo large deformations typically exploit large internal rotations, and therefore, necessitate the incorporation of flexible hinges. In the mechanism limit, these metamaterials consist of rigid bodies connected by ideal hinges that deform at zero energy cost. However, fabrication of structures in this limit has remained elusive. Here, we  demonstrate that the fabrication and integration of textile hinges  provides a scalable platform for creating large structured metamaterials with mechanism-like
behaviors.  Further, leveraging recently introduced kinematic optimization tools, we demonstrate that textile hinges enable extreme shape-morphing responses, paving the way for the development of the next generation of mechanism-based metamaterials.
\end{abstract}

\maketitle

\section*{Keywords}
mechanical metamaterials, architected structures, fabrication, nonlinear deformations, linkages

\section{Introduction}

From robots and engines to bikes and watches, mechanisms consisting of rigid bodies interconnected by joints or links are essential components in many mechanical systems~\cite{Howell2001}. Among these systems are mechanical metamaterials -- structures with unconventional properties emerging from their internal architecture~\cite{Christensen2015,Bertoldi2017}. A notable example is the rotating-square mechanism~\cite{Grima2000}, which consists of rigid squares linked by freely-pivoting hinges. This mechanism has played a pivotal role in the development of mechanical metamaterials with negative Poisson's ratio \cite{Bertoldi2010, Grima2000}, shape-morphing capabilities~\cite{Cho2014,choi2019programming} as well as programmable non-linear static~\cite{Deng2022} and highly-nonlinear dynamic~\cite{Bolei2017, Raney2020} responses. In these metamaterials the ideal freely-pivoting hinges are replaced by flexible components. To uphold the kinematic properties of the mechanism, it therefore becomes crucial to realize flexible  elements capable of achieving significant rotations at low energy cost. This requirement is often met by locally decreasing the material thickness, resulting in so-called living hinges~\cite{Bertoldi2010,Cho2014,Rafsanjani2016,Deng2022}. Nonetheless, because of fabrication constraints, employing this approach typically results in hinges that possess significant stiffness, dampening the desired idealized mechanical response~\cite{coulais_characteristic_2018}.

Here, we present a robust strategy to fabricate and characterize elastic, low-stiffness hinges using thin textile ribbons, first introduced in ~\cite{Dudte2023}.  Firstly, we show that the resulting textile hinges have versatile and reproducible mechanical properties, surpassing the flexibility observed in traditional living hinges. Secondly, we fabricate and test large arrays of textile-hinged counter-rotating squares and demonstrate good mechanism-like behaviour (as expected for rigid bodies joined by ideal hinges).
Lastly, utilizing advanced kinematic design algorithms~\cite{choi2019programming,choi2021compact}, we leverage textile hinges to develop shape-morphing structures. These structures can transition between distinct shapes, each representing an energy minimum, and facilitate mechanism-like behaviour while opening avenues for designing metamaterials with extreme  deformation responses.

\section{Textile hinges}

\begin{figure*}[ht]
	\centering
	\includegraphics[width=\linewidth]{{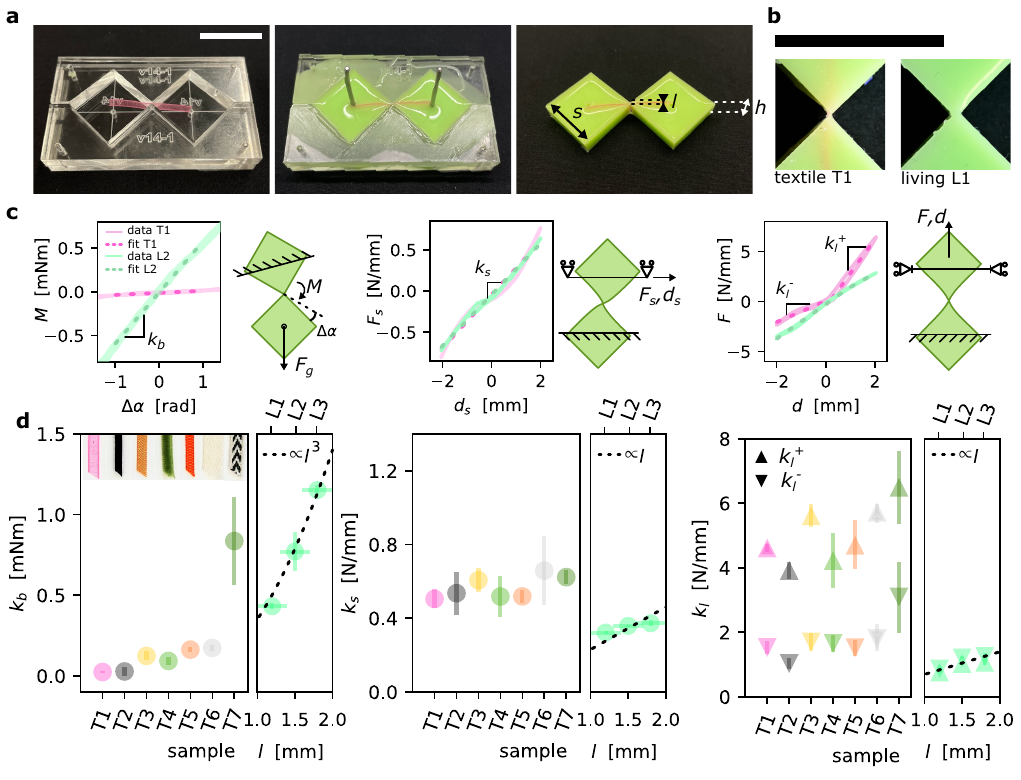}}
	\caption{
		\textbf{Mechanical response of textile and living hinges.}
		\textbf{a)} Snapshots taken during fabrication of a sample. 
		\textbf{b)} Snapshots of a textile hinge  $T1$ and a living hinge with $l=1.2$ mm. 
		\textbf{c)} Resistance to the hinge's three main deformation modes is characterized using mechanical tests. Typical results for textile hinges $T1$ and living hinges $L2$ are shown. Left: bending stiffness $k_b$ is obtained by measuring the moment $M$ and deflection angle $\Delta\alpha$ under gravity $F_g$.  Middle: shear stiffness $k_s$ relates shear displacement $d_s$ and force $F_s$. Right: longitudinal force $F$ and extension $d$ yield extension and compression stiffnesses $k_l^+$ and $k_l^-$.
        \textbf{d)} Resulting stiffness values for textile hinges $T1$-$T7$  and living hinges $L1$-$L3$. All data points are averaged over three distinct samples. Dash lines correspond to  the expected scalings from linear elasticity for living hinges, $k_b\propto l^3$ and $k_s, k_l \propto l$. 
        Scale bars: 20mm.
	}
	\label{def_fig1}
\end{figure*}

To realize flexible components that act as hinges with highly tunable behavior within the metamaterial, we augment living hinges with textile ribbons. We explore the potential of this strategy by first studying the mechanics of a variety of textile hinges and, second, comparing them to their living hinge counterparts. Towards this end, we first consider two square units connected at their vertices by a single hinge. As shown in Fig.~\ref{def_fig1}a,  specimens were manufactured by casting silicone rubber (Zhermack Elite Double, with a Shore-A hardness of 32 and an initial shear modulus of $G = 0.35$ MPa) within rigid molds. These molds were designed with two square voids, each featuring a side length of $s=20\pm 0.5$ mm and an out-of-plane thickness of $h=6.3\pm0.5$ mm, connected by a narrow channel. During the casting process, textile ribbons could be introduced into the channel before casting, resulting in specimens composed of two rubber squares connected by a slender textile ligament. Alternatively, the channel could be left open and filled solely with rubber, leading to an elastomeric living hinge with a width denoted as $l$ (see Supporting Information for additional details).

For this study, we have investigated seven distinct textile ribbons, each possessing unique characteristics (see Table S1 for detailed specifications). These textile ribbons include the following: ($T1$) a plain-weave polyester organza, ($T2$) suede, which combines a plain-weave polyester backing with fused polyester nap, ($T3$) a satin-weave polyester, ($T4$) a plain-weave nylon backing with a velvet nap, ($T5$) a plain-weave polyester grosgrain, ($T6$) a twill-weave cotton, and ($T7$) a twill-woven blend of cotton and polyester. Notably, all textiles, except for $T2$, have edges that are reinforced with chain stitching. Further, we have also considered  three living hinges with $l=1.2$, $1.5$, and $1.8 \pm 0.2$ mm, denoted as $L1$, $L2$, and $L3$, respectively. For visual reference, we have included representative close-up images of these hinges in both Fig.~\ref{def_fig1}b and Figure S3.

To characterize the  hinges, we test their response under imposed bending, stretching, and  shearing. These three modes of deformation have been demonstrated to be sufficient  to predict the behavior of arrays of solid units interconnected by slender hinges~\cite{deng2020,coulais_characteristic_2018}. The behavior upon bending was characterized by monitoring the angular deflection of the hinges under the influence of gravity and subsequently calculating the corresponding torque. To evaluate the response to stretching and shearing, we applied a displacement to one of the squares and recorded the resulting reaction force (see Supporting Information for experimental details). 

As shown in Fig.~\ref{def_fig1}c for squares with $s=20\pm0.5$ mm and $h=6.3\pm0.5$ mm,  upon both bending and shearing the hinges exhibit close to linear behaviour. Therefore, they can  be characterized by a linear  bending stiffness $k_b$ (which relates applied moment $M$ to angular deflection $\Delta \alpha$ via $M=k_b\Delta\alpha$) and a linear shear stiffness $k_s$ (which relates the shear force $F_s$ and the shear displacement $d_s$ via $F_s=k_sd_s$). Regarding stretching, the living hinges demonstrate nearly linear behavior, while the textile hinges exhibit bilinear behavior, with a more compliant response upon compression attributable to ribbon buckling. To describe this response we introduce both a tensile and a compressive axial stiffness $k_l^+$ and $k_l^-$, which relate the longitudinal force $F$ to the  displacement $d$ via $F=k_l^+d$ for $d>0$ and $F=k_l^-d$ for $d<0$. All measured stiffness values for squares with $s=20\pm0.5$ mm and $h=6.3\pm0.5$ mm, obtained via linear regression, are reported in Fig.~\ref{def_fig1}d and Table S2. Our results show that stiffness values for living  hinges scale as expected with hinge width $l$ within error margins, yielding  $k_b \propto l^3$, $k_s\propto l$, and $k_l \propto l$ (dashed lines in Fig.~\ref{def_fig1}d). Measured values lie in the range $k_b \in [0.4, 1.2]$ mNm, $k_s \approx 0.4$ N/mm, and $k_l \approx  1$ N/mm. Textile hinges exhibit comparable compressive stiffness values ($k_l^- \approx 1.5$ N/mm) to living hinges, but significantly greater tensile ($k_l^+\approx5$ N/mm) and shearing stiffness ($k_s\approx 0.6$ N/mm). Importantly, the choice of textile strongly affects the hinges' bending stiffness. Specifically, the bending stiffness of textile hinges varies over almost two orders of magnitude ($k_b\in[0.02, 1.2]$ mNm), matching the thickest living hinge  and ranging far below the thinnest living hinge.

Since stretching and shear introduce deformations that compete with the bending-dominated purely rotating mode of a mechanism, the mechanism-like quality of a hinge can be characterized by using two dimensionless parameters~\cite{coulais_characteristic_2018}-- a shear parameter $\alpha$ and a stretch parameter $\beta$:
\begin{equation}\label{eq:alphabeta}
\alpha = \frac{s^2k_s}{4k_b},\;\;\;\;\;\;\beta =\frac{ s^2 k_l}{4k_b}~.
\end{equation}
A hinge  approaches mechanism-like behavior  as the hinge width vanishes, i.e. as $\alpha \rightarrow \infty$ and $\beta\rightarrow \infty$. Using the average measured stiffness values presented in Fig.~\ref{def_fig1}d, we find that our living hinges have $\alpha \in [30, 75]$ and $\beta \in [100,200]$. By contrast, textile hinges exhibit $\alpha \in [70, 2200]$ and $\beta \in [300, 19000]$, indicating that textile hinges can more closely approach mechanism-like behavior.

\section{Textile-hinged metamaterials}

\begin{figure*}[ht]
	\centering
	\includegraphics[width=\linewidth]{{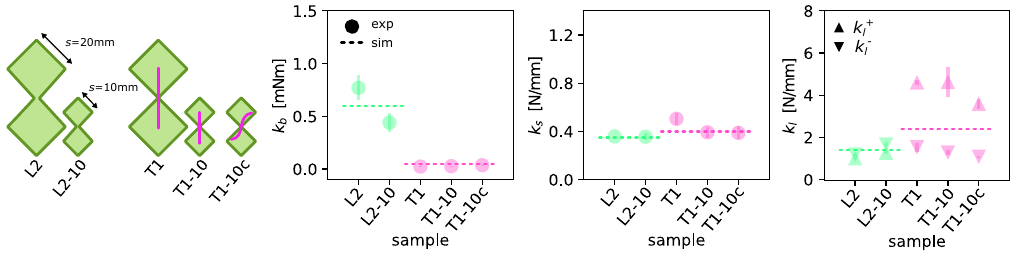}}
	\caption{
		\textbf{Effect of square size and ribbon orientation on hinge properties.}
         The mechanical properties of two-square samples with edge lengths $s = 20$ mm and $s = 10$ mm, connected either by living hinges $L2$ or textile hinges $T1$ (straight or curved), are explored. The sample thickness is kept constant at $h = 6.3 \pm 0.5$ mm.   Left to right: measured bending, shearing, and axial stiffness values for the samples.  Dashed lines correspond to stiffness values used for simulations in Fig.~\ref{def_fig2} and Fig.~\ref{def_fig3}.
	}
	\label{def_fig1_hingevariations}
\end{figure*}

\begin{figure*}
	\centering
	\includegraphics[width=\linewidth]{{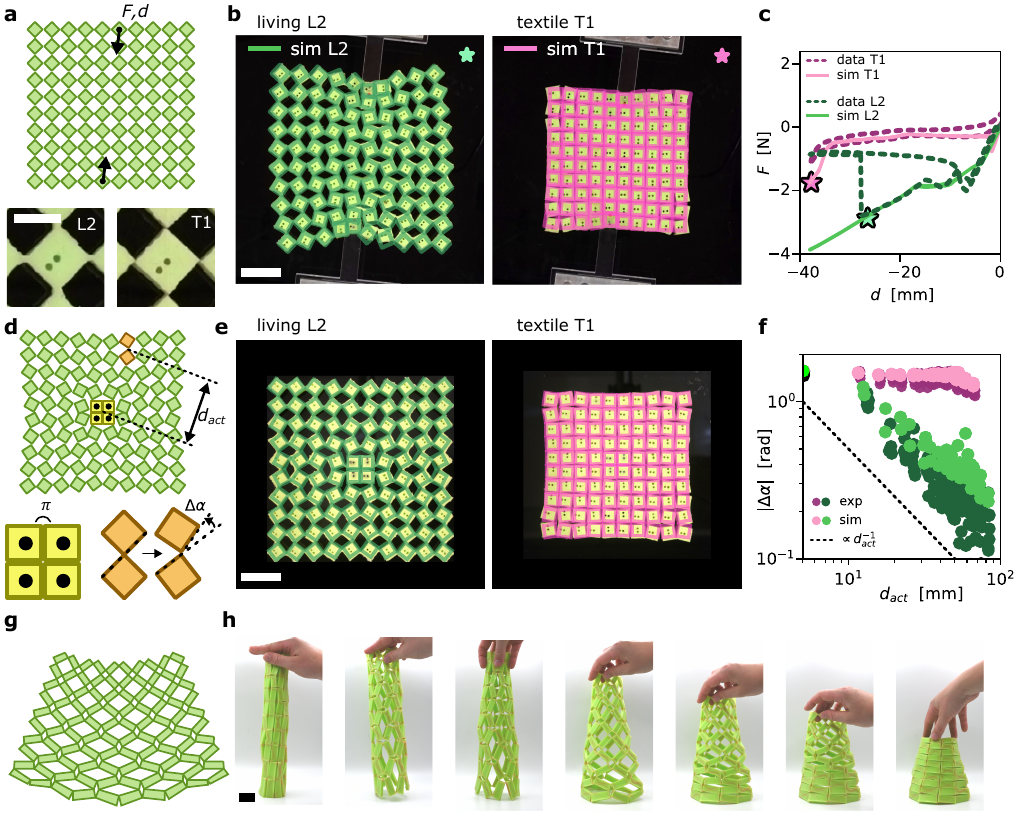}}
	\caption{
		\textbf{Textile-hinged metamaterials.}
       \textbf{a)} A $10\times10$ array of hinged squares with $s=10$mm and $h=6.3\pm0.5$mm is actuated under point loading (black arrows). Zoom-in of  living hinges $L2$ and textile hinges  $T1$ are shown (insets). Scale bar: 10mm.
        \textbf{b)} Experimental snapshots of the metamaterial with (left) living hinges $L2$ at an applied  displacement  $d=-27$mm (just prior to out-of-plane buckling) and  (right) textile hinges $T1$ at  $d=-38$mm. Numerical predictions are overlaid to experimental snapshots, showing an excellent agreement.
        \textbf{c)} Experimentally measured and simulated force $F$ and displacement $d$ for the two metamaterials.  Star markers correspond to snapshots in \textbf{b}.
        \textbf{d)} A $10\times10$ array of hinged squares with $s=10$mm and $h=6.3\pm0.5$ mm is loaded by rotating the central blocks (yellow squares)  by $\pi/4$ rad and pinning them.   
        \textbf{e)} Experimental snapshots of the metamaterial with (left) living hinges $L2$  and  (right) textile hinges $T1$. Numerical predictions are overlaid to experimental snapshots.
        \textbf{f)} Deformation decay is quantified by plotting $|\Delta\alpha|$ as a function of actuation distance $d_{act}$. Experimental measurements  and numerical predictions are shown for both samples   on a log-log scale to compare to the power-law decay $d_{act}^-1$ expected from a linear-elastic solid under point loading. Living hinges result in a material-like response with power-law decay, while textile hinges show mechanism-like behaviour with limited decay at underconstrained corners.
        \textbf{g)} Recently proposed mechanism-based design that forms a tubular structure capable of morphing
between a cylinder and a cone~\cite{choi2021compact}.
        \textbf{h)} Experimental snapshot of this shape-morphing design. The metamaterial is realized using rubbery blocks connected by textile hinges $T1$. 
        Scale bars \textbf{b}-\textbf{h}: 30mm.
	}
	\label{def_fig2}
\end{figure*}

The results of Fig.~\ref{def_fig1} demonstrate that textile hinges enable substantial rotations at a remarkably low energy cost. Consequently, when incorporated into a metamaterial, they are expected to generate a response closely resembling that of a  mechanism with kinematics primarily determined by geometry, rather than elasticity. To confirm this mechanism-like behaviour, we realized a $10\times10$ square-mechanism structure consisting of rubber squares with an edge length of $s=10$ mm and thickness $h=6.3\pm0.5$ mm, connected at their vertices either by textile hinges $T1$ or living hinges $L2$. Note  that the response of the hinges is not strongly affected by the size of the squares (Fig.~\ref{def_fig1_hingevariations}). However, to prevent ribbon crossing and ensure strong adhesion between the rubber and textile, the ribbons were woven into the molds along diagonal curved paths (see Fig. S2). The orientation of the textile ribbons across the hinge affects the axial tensile stiffness (Fig.~\ref{def_fig1_hingevariations}). Specifically, oblique or curved ribbons result in slightly lower values of $k_l^+$. This is likely due to the contribution of softer shearing in curved or oblique ribbons under sample extension, reducing the overall stiffness.

The metamaterial samples were fabricated by upscaling the previous molding method from 2 to 100 hinged squares (see Supporting Information for fabrication details). Remarkably, this upscaling did not present any challenges for textile hinges. In contrast, several living hinges experienced breakage during the demolding stage (see Fig. S4). In terms of robustness, tensile fracture testing on small square arrays shows a similar displacement-at-break for living and textile hinges $L2$ and $T1$. However, peak force is twice as large for textile hinges (see Fig. S6), indicating improved robustness under heavy loads.

We first study the mechanical response of the $10\times10$ square metamaterial under point loading, wherein a compressive displacement $d$ is applied to two squares located at opposing boundaries (Fig.~\ref{def_fig2}a). In the mechanism limit, this actuation is expected to lead to a complete collapse of the structure at zero force due to counter-rotations of neighbouring squares. As shown in Fig.~\ref{def_fig2}b, we find that in the sample with textile hinges, all squares indeed rotate by 45 degrees, taking it to its fully closed configuration at low forces ($F > -1$N, see Fig.~\ref{def_fig2}c). By contrast, in the sample with living hinges  only a few squares in the vicinity of the loading points undergo substantial rotation, before the structure buckles out-of-plane at $d=-7.6$mm and $F \approx -2$N.

Finally, it is important to note that these experimental observations are well captured by a simple model comprising  rigid units connected at their vertices by a combination of  bending, axial, and shear springs~\cite{deng2020,coulais_characteristic_2018,bordiga_2024}  with stiffness that match the experimentally measured values (i.e. $(k_b, k_s, k_l) = (0.05\,\mathrm{mNm},\, 0.4\,\mathrm{N/mm},\, 2.4\, \mathrm{N/mm})$ for hinges T1 and $(0.6\,\mathrm{mNm},\, 0.35\,\mathrm{N/mm},\, 1.4\, \mathrm{N/mm})$ for hinges L2; see Supporting Information for details).

To better understand how the hinges impact the metamaterials' deformation transmission capabilities, we carry out an additional experiment in which we rotate the four central blocks of each sample by 45$^\circ$ (Fig.~\ref{def_fig2}d). We quantify transmission capability by measuring the change in rotation angle $\Delta \alpha$ between each pair of connected blocks at distance $d_{act}$ from the sample's central actuation location. In Fig.~\ref{def_fig2}f we show the evolution of $\Delta \alpha$ as a function of $d_{act}$. The metamaterial with textile hinges behaves very similarly to a mechanism, for which $\Delta \alpha$ is expected to be constant across the sample,  with nearly lossless transmission except at the underconstrained corners of the array. By contrast, the transmission of deformation shows power-law decay in the metamaterial with living hinges---as expected for a linear-elastic structure under local actuation---with a decay length on the order of several square side lengths $s$. Finally, it is important to highlight that our model again accurately captures all experimental observations.

Since textile-hinged metamaterials closely approach the mechanism limit, they provide a robust platform for shape-morphing. While arrays of hinged squares offer limited shape changes, a recently proposed inverse-design framework based purely on kinematics has successfully identified tessellations of irregular quadrilaterals that can transform into a wide range of shapes~\cite{choi2019programming,choi2021compact,dang2021theorem}. Leveraging our fabrication strategy, which can be easily extended to units of arbitrary shapes, we realize one such tessellation that forms a tubular structure capable of morphing between a cylinder and a cone~\cite{choi2021compact} (Fig.~\ref{def_fig2}g-h). We fabricate the $10\times10$ design in its planar configuration (Fig.~\ref{def_fig2}g) using rubber quadrilaterals and textile hinges $T1$ (see Supporting Information for details). 
Wrapping the flat structure into a conical shape and bonding its opposing edges with the same rubbery material produces a three-dimensional conical form. As intended, this structure seamlessly transitions between the target cylindrical and compact conical configurations (Fig.~\ref{def_fig2}h and Supplementary Movie S1).

\section{Improving mechanism-like properties}

\begin{figure*}[ht]
	\centering
	\includegraphics[width=\linewidth]{{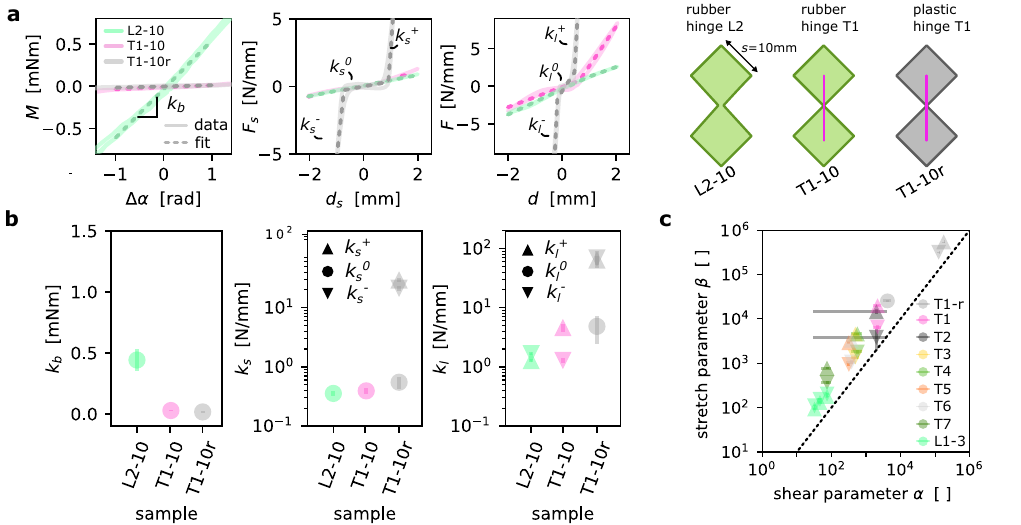}}
	\caption{
		\textbf{Effect of stiffness of the rotating units.}
        \textbf{a)} The mechanical properties of
two-square samples with edge lengths $s = 10$ mm made either of soft rubber or rigid plastic and connected either by living hinges $L2$ or textile hinges $T1$ are explored (see schematics, right).  Representative moment-angle, shear force-displacement, and axial force-displacement curves are shown  for all samples. 
        \textbf{b)} Left to right: measured bending, shearing, and axial stiffness values  for all samples. 
        \textbf{c)} Average dimensionless shear and stretch parameters $\alpha$, $\beta$ defined in Eq.~\ref{eq:alphabeta}  for   two-square samples with $s=20$ mm and $h=6.3\pm0.5$ mm connected by both textile and living hinges. 
	}
	\label{def_fig1bis}
\end{figure*}

As quantified by the dimensionless shear and stretch parameters $\alpha$ and $\beta$, textile hinges improve mechanism-like behaviour by up to two orders of magnitude when compared to living hinges. Further improvement can be achieved by increasing the stiffness of the matrix material used for the quadrilateral units. To demonstrate this, we compared the mechanical responses of two-square samples cast from soft silicone rubber (Zhermack Elite Double, shear modulus $G=0.35$ MPa) and rigid polyurethane plastic (Smooth-On Smooth-Cast 300, 
$G=2.9$ MPa). Samples were tested as shown in Fig.~\ref{def_fig1}c to measure their bending, shearing, and stretching responses. The recorded moment-angle and force-displacement curves are presented in Fig.~\ref{def_fig1bis}a, while the extracted stiffness values are shown in Fig.~\ref{def_fig1bis}b.
The results indicate that $k_b$ is primarily influenced by hinge type (living or textile) and only minimally by the stiffness of the blocks. By contrast, the shearing and longitudinal behavior of rigid plastic blocks connected by a textile hinge differs from that of rubber blocks connected by living or textile hinges. Under both shear and axial deformations, the force response exhibits three approximately linear regimes. For small displacements, the behavior is dominated by the flexible textile hinge, resulting in recorded forces similar to those observed for the rubbery squares and quantified by stiffnesses 
$k_s^0$ for shear and 
$k_l^0$ for longitudinal actuation. However, for sufficiently large applied displacements, the rigid response of the plastic blocks engages, resulting in higher stiffness values 
$k_s^-$, $k_s^+$ and $k_l^-$, $k_l^+$
  under negative and positive shearing and stretching. 
  
  Importantly, these higher stiffness values under shearing and longitudinal deformation lead to larger values of the dimensionless shear and stretch parameters 
$\alpha$ and $\beta$ under large deformations. As shown in Fig.~\ref{def_fig1bis}c, for textile hinges 
$T1$ embedded in blocks made of rigid plastic, 
$\alpha\in[120000,180000]$ and $\beta\in[310000, 540000]$, showing an improvement of around two orders of magnitude in mechanism-like behavior under large displacements compared to the same hinges connecting rubbery blocks ($\alpha\approx 2000$, $\beta\in[6000, 19000]$).

\begin{figure*}
	\centering
	\includegraphics[width=\linewidth]{{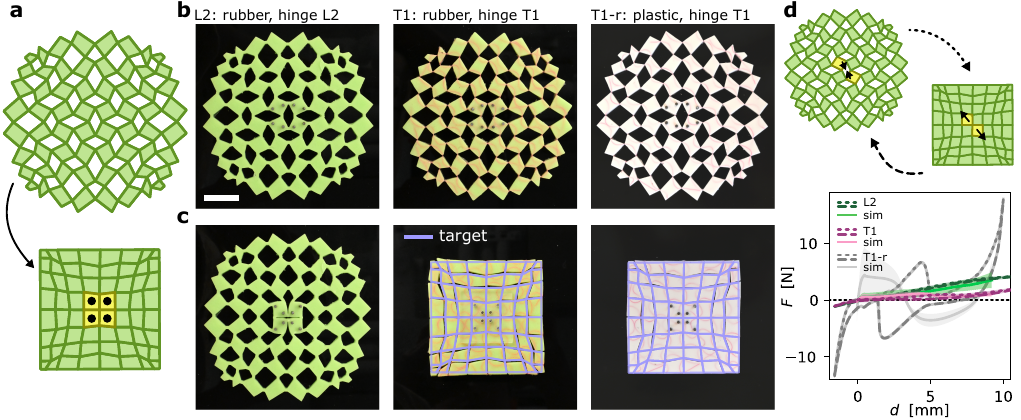}}
	\caption{
		\textbf{Multistable shape-morphing metamaterials.}
        \textbf{a)}  Previously identified kinematic design that transitions from an open circle to a compact square~\cite{choi2019programming}. 
        \textbf{b)} Snapshots of the samples in their open configuration. We fabricated three  samples     using rubbery material and living hinges $L2$ (left), rubbery material and textile hinges $T1$ (middle), and plastic material and textile hinges $T1$ (right). 
        \textbf{c)} Snapshots of the three samples  upon clamping their central four blocks into the compact rotated configuration. For the sample with textile hinges the target shape is overlaid to experimental snapshots, showing
an excellent agreement. 
        \textbf{d)} Additional tests were conducted where   two diagonally opposing central blocks where displaced as shown in the schematics. Experimental and simulated force-displacement curves of rubber and plastic metamaterials with hinges $L2$ and $T1$ (legend) show a good qualitative match. Bistability, signified by negative forces, is observed only for the sample with rigid plastic blocks ($T1-r)$. Simulations match experimental force magnitude and stability across samples.
        Scale bar: 20mm.
    }
	\label{def_fig3}
 \end{figure*}

 The improved mechanism-like behavior provided by stiffer blocks can be utilized to realize shape-morphing multistable structures. To demonstrate this, we considered a previously identified kinematic design that transitions from an open circle to a compact square (Fig.~\ref{def_fig3}a) \cite{choi2019programming}. In the limit of $\alpha,\,\beta\rightarrow\infty$, the design is predicted to be stable in its open and closed configurations \cite{choi2019programming}. We fabricated three samples: one with living hinges $L2$ connecting blocks of rubbery material, one with textile hinges $T1$ connecting blocks of rubbery material, and one with $T1$ hinges connecting blocks of stiffer plastic (Fig.~\ref{def_fig3}b). The samples were fabricated in their open circular configuration, and we tested their shape-morphing capability by clamping their central four blocks into the compact rotated configuration (Fig.~\ref{def_fig3}c).
 
The sample with living hinges exhibits deformations strongly localized near the clamped blocks (Fig.~\ref{def_fig3}c, left), confirming that metamaterials with living hinges do not generally support mechanism-like behaviors. Fabrication constraints limit their minimum width $l$, and consequently their shear and stretch factors $\alpha$ and $\beta$. As a result, computationally simple inverse-design strategies based purely on kinematics cannot be used; instead, design optimization must rely on a fully elastic model~\cite{bordiga_2024}, which is generally more computationally expensive (see Supporting Information for details).
In contrast, the textile-hinged samples almost completely collapse to the compact square target (overlaid lines in Fig.~\ref{def_fig3}c, center and right). However, while the textile-hinged sample with rubbery blocks immediately return to its open circular configuration when the loading is removed, the one with stiff plastic units retain its square configuration, indicating multistable behavior (see Supplementary Movie S2).

This multistable behavior is confirmed by mechanical tests  performed on the three samples by applying a displacement to two central blocks (Fig.~\ref{def_fig3}d, top). The resulting force-displacement curves reveal two distinct behaviour. For the samples with rubbery blocks, the sign of the force matches the actuation direction, and the samples immediately return to their open states as soon as the loading is removed. However, for the sample with stiff plastic blocks, the loading process is characterized by a snapping event where the structure suddenly transforms into the compact square configuration. This snapping event corresponds to the sharp drop in load at a displacement $d \approx 5$ mm. When the load is removed, the measured load drops below zero and the sample retains its square shape  (Fig.~\ref{def_fig3}d, bottom). Note that simulations using the numerical model introduced in Fig.~\ref{def_fig2} reproduce stability behaviour and measured force magnitudes across samples (continuous lines in Fig.~\ref{def_fig3}d - see Supporting Information for details). Thus, our findings demonstrate that the large values of $\alpha$ and $\beta$, achieved by introducing stiff blocks, make the structure sufficiently mechanism-like to exhibit multistability.
 
\section{Discussion and Conclusions}
In summary, we have demonstrated that the integration of textile hinges provides a versatile framework for creating metamaterials with  mechanism-like behaviour, bridging the gap between computationally based kinematic design strategies and existing fabrication methods for robust hinges. Additionally, we have shown that increasing the stiffness of the blocks can enhance this mechanism-like behavior, providing opportunities to control the energy landscape of the metamaterials and achieve multistability. While our current focus has centered on metamaterials based on the rotating-square mechanism, our methodology extends to  other families of metamaterials in which internal rotations play a dominant role, such as origami~\cite{Callens2018,Dieleman2020,Dudte2021,Zhai2021} and kirigami-inspired~\cite{Celli2018,Tao2023,Jin2024} designs. As such, the approach outlined in this study shows potential for integration into the upcoming wave of metamaterials~\cite{Chi2022,Dudte2023,Liu2024}, which harness programmable shape-morphing~\cite{Wang2023,Qiao2024} and multistability~\cite{Peng2024} to achieve functionality, paving the way for designing the next generation of intelligent structures and soft robots.

\section*{Acknowledgements} Research was supported by  NSF grant DMR-2011754 and Army Research Office grant W911NF-22-1-0219. 
Thanks to Adel Djellouli, Davood Farhadi, and Giada Risso for insightful discussions.

\section*{Conflict of interest} 
The authors declare no conflict of interest.

\section*{Supporting Information} 
Supporting Information is available from the Wiley Online Library or from the author.

\section*{Code availability}
Code underlying the findings in this work has been made available on GitHub at \href{https://github.com/bertoldi-collab/DifFlexMM}{github.com/bertoldi-collab/DifFlexMM} and \href{https://github.com/bertoldi-collab/BlockyMetamaterials/tree/TexMorpherSims}{github.com/bertoldi-collab/BlockyMetamaterials/tree/TexMorpherSims}.

\clearpage
\newpage


\title{Supporting Information for \textit{Textile hinges enable extreme properties of mechanical metamaterials}}

\author{A.~S.~Meeussen*}
\affiliation{School of Engineering and Applied Sciences, Harvard University, Cambridge,	Massachusetts 02138, USA}
\author{G.~Bordiga}
\affiliation{School of Engineering and Applied Sciences, Harvard University, Cambridge,	Massachusetts 02138, USA}
\author{A.~X.~Chang}
\affiliation{School of Engineering and Applied Sciences, Harvard University, Cambridge,	Massachusetts 02138, USA}
\author{B.~Spoettling}
\affiliation{ETH Zuerich, Zuerich 8092, Switzerland}
\author{K.~P.~Becker}
\affiliation{School of Engineering and Applied Sciences, Harvard University, Cambridge,	Massachusetts 02139, USA}
\affiliation{Department of Mechanical Engineering, Massachusetts Institute of Technology, Cambridge, Massachusetts 02139, USA}
\author{L.~Mahadevan}
\affiliation{School of Engineering and Applied Sciences, Harvard University, Cambridge,	Massachusetts 02138, USA}
\author{K.~Bertoldi}
\affiliation{School of Engineering and Applied Sciences, Harvard University, Cambridge,	Massachusetts 02138, USA}

\maketitle

\section{Fabrication}

 \begin{figure*}[ht]
    \begin{center}      
    \includegraphics[width=\linewidth]{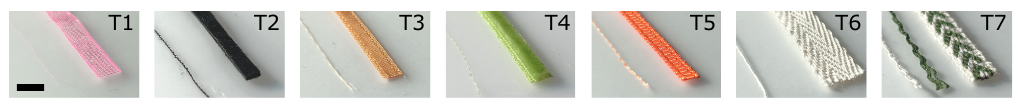}   
        \caption{
        \textbf{Textile ribbons used to create metamaterial hinges.} 
            Each panel shows one of the fabric types, labelled T1 through T7. A representative ribbon section and weft fiber is shown for each type. Details are described in Table~\ref{table:SI_ribbontypes}. Scale bar: 5mm.
            }
        \label{fig:SI_ribbontypes}
        \vspace{-15pt}
    \end{center}
\end{figure*}

\begin{table}[ht]
\centering
 \begin{tabular}{|c c c c c c c|} 
 \hline
 ID & width (mm) & material & weave & weft fibers & selvedge & link\\ [0.5ex] 
 \hline\hline
 T1 & 4.0$\pm$0.2 & polyester & plain &  26 & chain stitch & tinyurl.com/dhorganza\\ 
 T2 & 2.4$\pm$0.2  & polyester & plain, fused nap & 9 & fused & tinyurl.com/dhsuede\\
 T3 & 3.6$\pm$0.2 & polyester & satin & 34 & chain stitch & tinyurl.com/dhsatin\\
 T4 & 3.4$\pm$0.2 & nylon & plain, velvet nap & 28 & chain stitch & tinyurl.com/dhgreen\\
 T5 & 2.8$\pm$0.2 & polyester & plain & 35 & chain stitch & tinyurl.com/dhorange\\ [1ex] 
 T6 & 7.4$\pm$0.2 & cotton & twill & 26 & chain stitch & tinyurl.com/dhtwill\\ [1ex] 
 T7 & 5.4$\pm$0.2 & cotton, polyester & twill & 36 & chain stitch & tinyurl.com/dhchevron\\ [1ex] 
 \hline
 \end{tabular}
            \caption{
            \textbf{Details of textile ribbons as shown in Fig.~\ref{fig:SI_ribbontypes}.} 
                }
            \label{table:SI_ribbontypes}
\end{table}

Here, we describe the fabrication of samples used in this work. All samples were fabricated by molding either green two-compound platinum-cure polvinyl siloxane (PVS) rubber (Zhermack Elite Double 32, initial shear modulus $G=0.35$MPa) or white two-compound polyurethane (PU) liquid plastic (Smooth-On Smooth-Cast 300, initial shear modulus $2.9$MPa). The fabric ribbons used to create textile hinges, labelled T1 through T7 as shown in Fig.~\ref{fig:SI_ribbontypes}, are described in more detail in Table~\ref{table:SI_ribbontypes}. Commercial ribbons with a variety of widths, materials, weave patterns, and weft fiber density were chosen. The fabrics' widths vary between $2.4\pm 0.2$ mm and $7.4\pm0.2$ mm.

    \subsection{Two hinged squares} 

    \begin{figure*}[h]
        \begin{center}        
        \includegraphics[width=\linewidth]{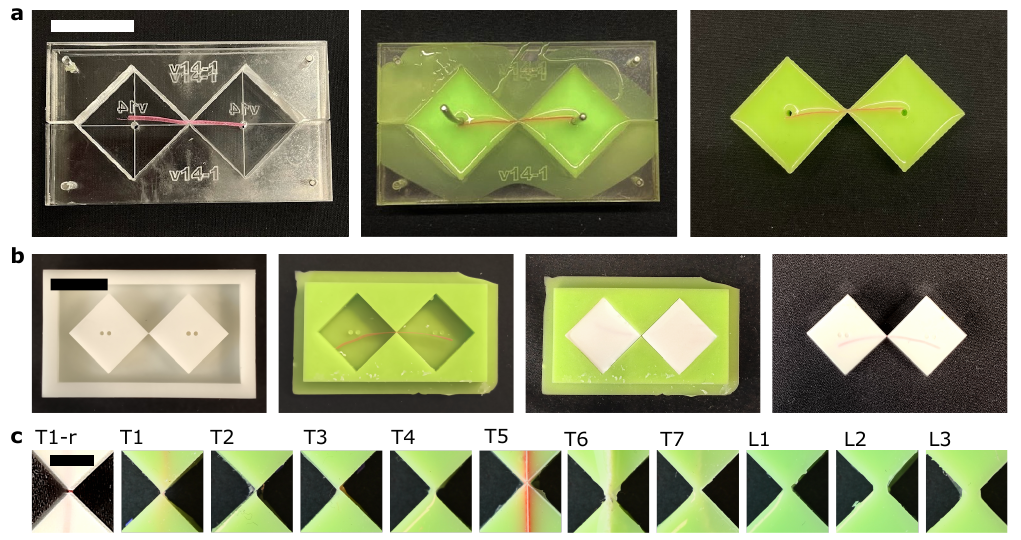}
            \caption{
            \textbf{Sample fabrication of two squares connected by a hinge.}
            \textbf{a)} Typical molding process for soft rubbery samples. Left to right: two sheets of laser-cut $6.3$mm-thick acrylic form a mold for two squares (side length $s=20$mm as shown, or $s=10$mm) connected by an open channel. A fabric strip may be introduced in the channel and rubber is cast in the mold. Unmolding and post-processing produces two squares connected by a textile or living hinge.
             \textbf{b)} Molding process for rigid plastic samples. Left to right: a 3D-printed plastic mold is used to cast a soft rubber mold with square inclusions connected by a channel. The  mold is filled with a textile hinge and liquid plastic, resulting in stiff $6.3$mm-thick squares connected by a textile hinge.
             \textbf{c)} Representative photos of textile hinge T1 connecting rigid plastic squares (T1-r), and textile (T1-T7) and living hinges (L1-L3) connecting soft rubber squares. Textiles are labelled as in Fig.~S1. Samples L1, L2, and L3 were designed to have progressively thicker hinges, measured to be $1.2\pm0.2$, $1.5\pm0.2$, and $1.8\pm0.2$ mm respectively.
            Scale bar: $5$mm.
            }
            \label{fig:SI_fabrication}
            \vspace{-15pt}
        \end{center}
    \end{figure*}
    
    As shown in Fig.~\ref{fig:SI_fabrication}a, two hinged rubber squares are fabricated by creating a mold from laser-cut $6.3$ mm-thick acrylic plate. A top plate, forming a void in the shape of two squares joined by a slim channel, is secured on a base plate with dowel pins. Dowel pins are placed in the center of each square to create clamping points in the sample for future mechanical testing. A fabric strip may be placed straight or curved in the squares' connecting channel to form a textile hinge (samples T1-T7); alternatively, the channel may be left open to create a living hinge (samples L1-L3). Liquid PVS rubber is cast in the resulting mold. After scraping away excess rubber and curing for at least 20 minutes, the hardened rubber is removed from the mold and post-processed to remove any flashing. Fig.~\ref{fig:SI_fabrication}b shows the fabrication procedure for hinged plastic squares. A 3D-printed structure (created with a Formlabs Form 3B stereolithography printer with Formlabs Rigid 4000 resin) is used to produce a PVS rubber mold with two square voids connected by a channel. Slender pillars in the square voids create clamping points for mechanical testing. Placing textile ribbon T1 in the channel, casting liquid PU in the mold, curing for 20 minutes, and post-processing to remove flashing results in a rigid plastic sample (denoted T1-r). Resulting samples have a height of $h=6.3\pm0.5$mm and square side lengths $s=10$ and $s=20$mm. Representative close-ups of all hinge and matrix types explored in this work are shown in Fig.~\ref{fig:SI_fabrication}c.\\

    \subsection{Arrays of hinged quadrilaterals}

    \begin{figure*}[h]
        \begin{center}        
        \includegraphics[width=\linewidth]{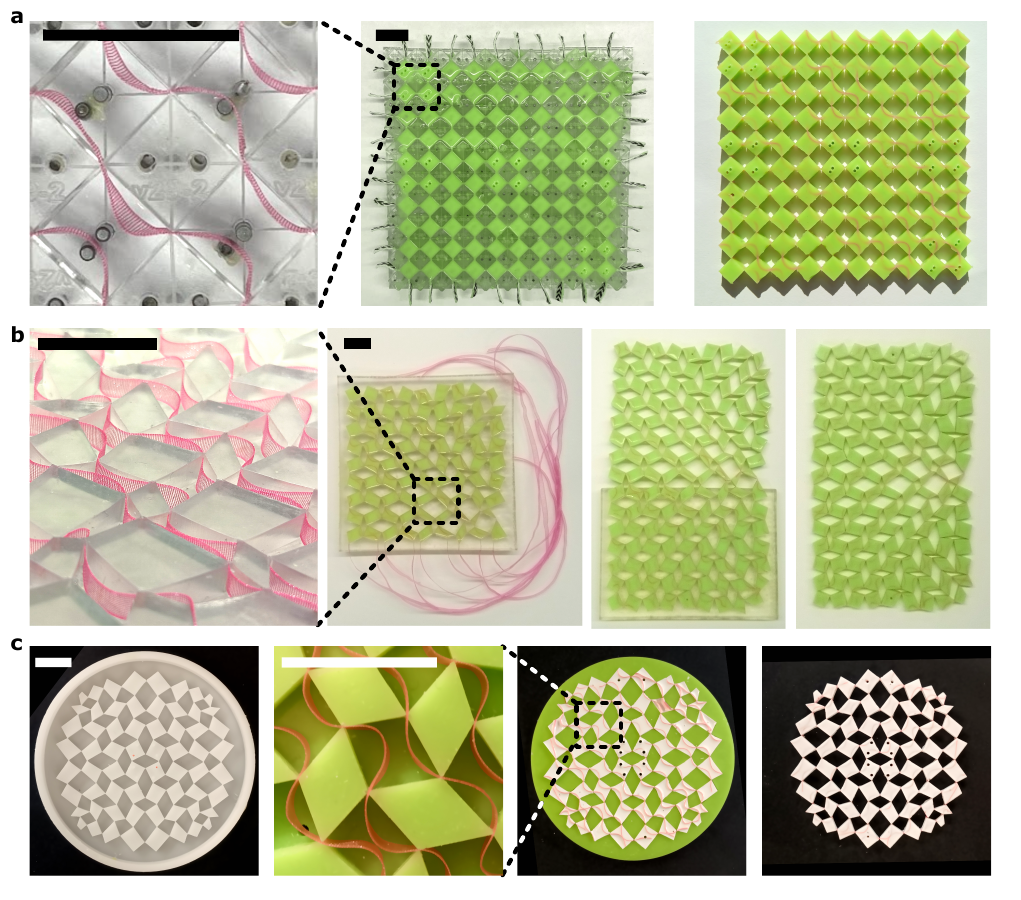}
            \caption{
            \textbf{Fabrication of large arrays of quadrilaterals.}
            \textbf{a)} Molding arrays of rubber squares. Left to right: $6.3$mm-thick laser-cut acrylic squares, attached to a base plate, form a mold for a 10-by-10 array of connected square voids with side length $s=10$mm. Fabric may be woven through the channels as shown. Casting, curing, and post-processing yields the final sample.
            \textbf{b)} Creating arrays of rubber quadrilaterals. Left to right: a mold with irregular quadrilateral voids (typical side length $s\approx 10$mm) is 3D-printed out of flexible material. Channels between voids may be woven through with long textile strips; rubber casting results in an irregular array.  Multi-part molding for large structures is achieved by weaving excess fabric through a second mold and repeating the procedure, resulting in samples such as the 9-by-15 array shown. Sample height is $4\pm0.5$mm.
            \textbf{c)} Arrays of rigid plastic units. Left to right: a 3D-printed form is used to create a soft rubber mold. Weaving textiles through the channels and filling voids ($s\approx 10$mm) with liquid plastic results in the final structure with height $4\pm0.5$mm.
            Scale bars throughout: $20$ mm.
            }
            \label{fig:SI_fabrication_arrays}
            \vspace{-15pt}
        \end{center}
    \end{figure*} 
    
    We distinguish three types of square lattices consisting of quadrilaterals: rubber squares (Fig.~\ref{fig:SI_fabrication_arrays}a); rubber quadrilaterals (Fig.~\ref{fig:SI_fabrication_arrays}b); and plastic quadrilaterals (Fig.~\ref{fig:SI_fabrication_arrays}c). 
    
    First, arrays of hinged rubber squares are molded in similar fashion to two hinged rubber squares (recall Fig.~\ref{fig:SI_fabrication}a). Laser-cut acrylic squares are fastened to an acrylic base plate with dowel pins; additional dowel pins are added to some squares to facilitate future testing. Fabric strips may be woven diagonally through the square's hinge channels. Curved diagonal textile paths are chosen to prevent ribbon crossing and ensure good adhesion between hinge and square material. Alternatively, the channels may be left open to create living-hinge arrays. Liquid PVS rubber is cast in the resulting mold. After scraping away excess rubber and curing for at least 20 minutes, the sample is removed from the mold and flashing is removed. Resulting arrays have square side lengths $s=10$mm and height $h=6.3\pm0.5$mm.
    
    Second, arrays of irregular rubber quadrilaterals are created by 3D-printing a flexible mold (using a Formlabs Form 3B stereolithography printer with Formlabs Elastic 50A resin) that contains quadrilateral voids connected by slim channels. Slender pillars are printed in select quadrilaterals to create clamping points for mechanical testing. Due to printer size constraints, the mold is printed in two parts and molding is subsequently done in a two-step process. First, fabric strips are woven diagonally through the channels of one mold, leaving excess fabric to connect to the second mold. Liquid PVS rubber is cast in the mold and post-processed as discussed above, resulting in a partial sample. The partial sample is then aligned with the second mold and its excess fabric is woven through the second mold's channels. Casting rubber in the second mold and post-processing as before results in a seamlessly connected array of irregular rubber quadrilaterals. The resulting samples have typical quadrilateral side length $s\approx10$mm and height $h=4\pm0.5$mm.        
    
    Third and last, lattices of plastic quadrilaterals are created similarly to hinged squares (see Fig.~\ref{fig:SI_fabrication}b). A primary 3D-printed mold is used to create a secondary rubber mold with suitable quadrilateral voids and connecting channels. Dowel pins are inserted into some voids to facilitate clamping during testing. Fabric ribbons of type T1 are woven diagonally through the channels before casting liquid PU. Curing and post-processing as described above results in textile-hinged samples with typical quadrilateral side length $s\approx10$mm and height $h=4\pm0.5$mm.        

    \subsection{Three-dimensional tubular array}
       
    \begin{figure*}[h]
        \begin{center}        
        \includegraphics[width=\linewidth]{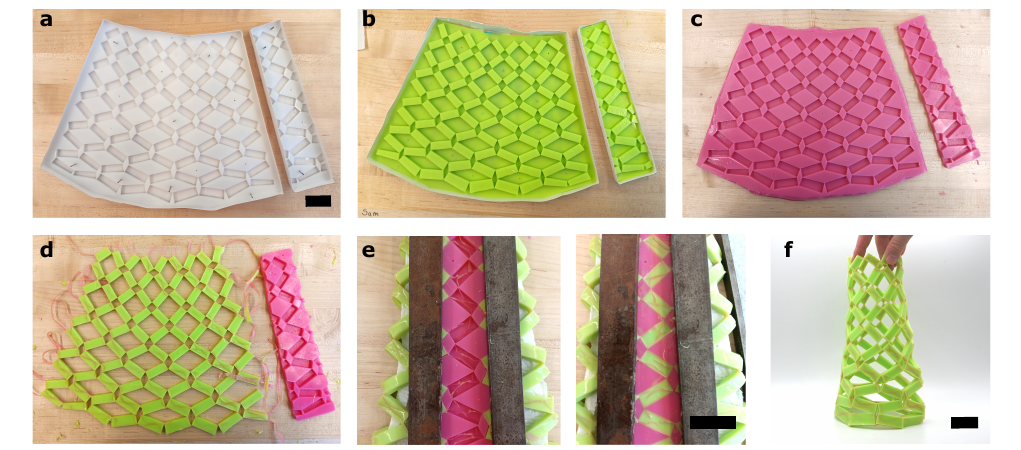}
            \caption{
            \textbf{Five-step fabrication of a three-dimensional tubular array of quadrilaterals.}
            \textbf{a)} The first mold is an array of irregular squares, laser-cut from $6.3$mm-thick acrylic and attached to a base plate.
            \textbf{b)} Soft PVS rubber is cast onto the acrylic array, forming the second mold.
            \textbf{c)} Rubber-on-rubber casting yields the third and final mold (shown in pink).
            \textbf{d)} The mold is woven through with textile hinges $T1$ and filled with soft PVS rubber, resulting in a flat quadrilateral array with excess edge fabric.
            \textbf{e)} The array is wrapped into a cylinder and its edges and excess textile are aligned in a partial mold (pink). Casting rubber in the partial mold completes the structure.
            \textbf{f)} Resulting conical structure.
            Multi-step molding ensures minimal rubber intrusion at the hinge, ensuring low rotational stiffness $k_b$ and good mechanism-like behaviour.
            Scale bars throughout: $30$ mm.
            }
            \label{fig:SI_fabrication_tubular}
            \vspace{-15pt}
        \end{center}
    \end{figure*}

    The three-dimensional array~\cite{choi2021compact} shown in Fig.3g-h, consisting of rubber quadrilaterals of typical side length $s\approx20$mm and thickness $h=6.3\pm0.5$mm connected by textile hinges T1, is fabricated in five steps using three molds (see Fig.~\ref{fig:SI_fabrication_tubular}). The first mold is created by laser-cutting quadrilateral voids into an acrylic sheet, which is then fastened on a base plate using dowel pins (Fig.~\ref{fig:SI_fabrication_tubular}a). Casting PVS rubber in the acrylic form yields the second mold (Fig.~\ref{fig:SI_fabrication_tubular}b). Creating a negative imprint by again casting PVS rubber (Fig.~\ref{fig:SI_fabrication_tubular}c) creates the third and final mold. As before, textile T1 is woven diagonally through the channels of the final mold, and excess textile is left at the edges. Casting, scraping, and curing liquid PVS rubber results in a flat array (Fig.~\ref{fig:SI_fabrication_tubular}d). Note that, to prevent adhesion between cured PVS rubbers, sufficient mold release agent must be applied, and structures must be separated as soon as curing allows. Finally, the flat array is wrapped into a three-dimensional conical shape. Its edges and excess textile are arranged into a partial mold (Fig.~\ref{fig:SI_fabrication_tubular}e) in which liquid PVS rubber is cast to complete the structure. The resulting three-dimensional tubular material, which has been post-processed to remove flashing, is shown in Fig.~\ref{fig:SI_fabrication_tubular}f. Note that this tubular metamaterial may also be fabricated using the technique illustrated in Fig.~\ref{fig:SI_fabrication_arrays}b.

\clearpage
\newpage

\section{Testing}
Here, we describe the testing procedures  used in this work. First, we discuss how we measure the effective behaviour of two squares connected by a slender hinge. Second, we show the protocols used to characterize mechanical properties of larger arrays of connected quadrilaterals. Third, fracture properties of square arrays with varying hinge types are explored.\\

     \subsection{Hinge characterization}     
\begin{figure}[h]
        \begin{center}        
        \includegraphics[width=0.5\linewidth]{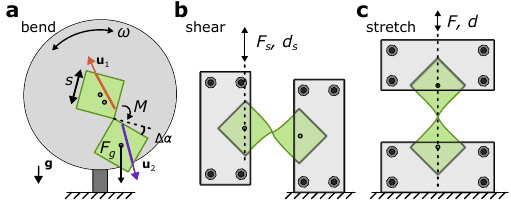}
            \caption{
            \textbf{Schematic testing setups to characterize hinge properties.}
            \textbf{a)} Hinge bending. A two-square sample of side length $s$ is pinned at one square to a vertical stage rotating back and forth at speed $\omega\approx 0.15$rad/s. Gravitational force $F_g$ working on the center of mass of the free square results in torque $M$ and angular hinge deflection $\Delta\alpha$, which are calculated by tracking square orientations $\uvec{u}_1$ and $\uvec{u}_2$ from a video recording using two colour-contrasting strips glued along each square's diagonal (coloured lines). 
            \textbf{b)} Hinge shearing. Each square is clamped between two plates with a partial square cutout fastened with nuts and bolts.
             One clamp is stationary, while the other is translated orthogonal to the sample's long axis as shear force $F_s$ and displacement $d_s$ are recorded. 
             \textbf{c)} Hinge stretching. Clamping is identical to the shearing configuration, but the free clamp is translated along the long axis of the two-square sample while axial force $F$ and displacement $d$ are recorded.
            }
            \label{fig:SI_testing_hinges}
        \end{center}
    \end{figure}

     The mechanical behaviour of two  squares connected by a soft hinge is dominated by the hinge's flexural properties under bending, shear, and axial deflection. Notably, these properties can be well captured by introducing three stiffness values \cite{coulais_characteristic_2018,deng2020}: bending stiffness $k_b$, relating applied moment $M$ to angular deflection $\Delta \alpha$ via $M=k_b\Delta\alpha$;  shear stiffness $k_s$ which relates shear force $F_s$ and shear displacement $d_s$ via $F_s=k_sd_s$; and axial stiffness $k_l$, relating axial force $F$ and elongation $d$ via $F=k_ld$. For this work, we distinguish axial compressive and extensile stiffnesses $k_l^-$ and $k_l^+$ for soft rubber squares, as well as small- and large-displacement shear and axial stiffnesses $k_s^0, k_s^{\pm}$ and $k_l^0, k_l^{\pm}$ for rigid plastic squares. The schematic testing setups used to determine $k_b$, $k_s$, and $k_l$ are shown in Fig.~\ref{fig:SI_testing_hinges}a--c.\\
     
     \emph{Bending stiffness:} Bending stiffness values for all two-square samples in this work were measured by tracking their hinges' angular deflection under gravity (Fig.~\ref{fig:SI_testing_hinges}a). To achieve this, one of the squares  is pinned vertically to a rotating stage, so that it moves with the stage while the other moves freely. A weight is then pinned to the center of the free square, bringing the free square's weight to a total mass $m$ , resulting in a gravitational force $F_g=m\,g$ at its center of mass along the vertical direction ($g=9.81$ m/s$^2$ denoting the gravitational acceleration). The rotation stage is driven via a pinion engaged with a rack that is attached to a translation stage. This produces a back-and-forth rotation at low angular speed $\omega=0.15\pm0.3$ rad/s and results in deflection of the sample. A video recording of the sample is used to track the squares' orientations $\uvec{u}_1$ and $\uvec{u}_2$, via two strips of colour-contrasting plastic glued along each square's diagonal. Their relative angular deflection is then calculated via 
     \begin{equation}
        \Delta \alpha = \pi - \arccos(\uvec{u}_1\cdot \uvec{u}_2),
     \end{equation} 
     and the corresponding  torque can be obtained as
     \begin{equation}
       M= F_g \frac{1}{\sqrt{2}}l \uvec{u}_2 \times \uvec{g}.  
     \end{equation}\\
     
     \emph{Axial and shear stiffness:} To obtain axial and shear stiffness values for all the samples  with square side length $s=20$ mm we used a universal testing machine (Instron 5969, outfitted with a 2530-series $50$-N load cell). By contrast, axial and shear stiffness values for samples with square side length $s=10$ mm and hinges consisting either of  textile ribbon T1 or elastomeric thin ligament L2 with a width of $l=1.5\pm0.3$ mm were measured using a pair of translation stages (Thorlabs LTS300 and LTS150) outfitted with a load cell (Futek LSB200, capacity $50$ lb) and controlled with a custom Python library. Samples were actuated at $0.5$mm/s throughout to ensure quasistatic loading. Schematic illustrations of the clamping and loading conditions are shown in Fig.~\ref{fig:SI_testing_hinges}b,c.\\
    
      \emph{Results:} Stiffness values were obtained from all torque-angle and force-displacement measurements by fitting with linear or piecewise linear functions (see main text, Fig.~1, and Fig.~4). Collected results, averaged over three instances per sample type, are presented in Table~\ref{table:SI_testing_hinges} (top). In addition, shear and stretch parameters $\alpha$ and $\beta$, which quantify the mechanism-like properties of hinges (see main text, Eq.~1), were calculated according to 
      \begin{equation}\label{eq:alphabeta}
        \alpha = \frac{s^2k_s}{4k_b},\;\;\;\;\;\;\beta =\frac{ s^2 k_l}{4k_b}~.
    \end{equation}
    Resulting values, also shown graphically in Fig.~4c, are reported in  Table~\ref{table:SI_testing_hinges} (bottom). Errors, quantified by standard deviations $\sigma$, were calculated using linear propagation of uncertainty according to
        \begin{align}\label{eq:alphabeta_error}
        \sigma_{\alpha} = \alpha \sqrt{(\sigma_{k_b}/k_b)^2 + (\sigma_{k_s}/k_s)^2},\;\;\;\;\;\;\
        \sigma_{\beta} = \beta \sqrt{(\sigma_{k_b}/k_b)^2 + (\sigma_{k_l}/k_l)^2}~.
        \end{align}
        
   \begin{table}[h!]
        \tiny
        \centering
        \begin{tabular}{|cccccccccccccccccccccc|}
        \hline
           & $k_b$ & & & $k_s^-$  & & &  $k_s^0$  & & & $k_s^+$  & & & $k_l^-$  & & & $k_l^0$  & & & $k_l^+$ &  &\\           
           &  [mNm] & & & [N/mm] & & &   [N/mm] & & & [N/mm] & & &  [N/mm] & & & [N/mm] & & & [N/mm] &  &\\
        ID & avg. & $\pm$ range & std. & avg. & $\pm$ range & std.& avg. & $\pm$ range & std.& avg. & $\pm$ range & std. & avg. & $\pm$ range & std. & avg. & $\pm$ range & std. & avg. & $\pm$ range & std.\\
        \hline
        \hline
        T1 & 0.023 & 0.007 & 0.005 & &   &   &  0.50 & 0.06 & 0.05 & &   &   & 1.53 & 0.24 & 0.20 &   &   & & 4.61 & 0.16 & 0.13  \\
        T2 & 0.027 & 0.027 & 0.026 & &   &   & 0.54 & 0.14 & 0.12 & &   &   & 1.03 & 0.21 & 0.17 &   &   & & 3.91 & 0.31 & 0.25  \\
        T3 & 0.12 & 0.03 & 0.03 & &   &   & 0.61 & 0.08 & 0.07 & &   &   & 1.70 & 0.34 & 0.27 &   &   & & 5.61 & 0.42 & 0.35 \\
        T4 & 0.09 & 0.03 & 0.02 & &   &   & 0.52 & 0.14 & 0.11 & &   &   & 1.65 & 0.32 & 0.27 &   &   & & 4.22 & 1.04 & 0.85  \\
        T5 & 0.16 & 0.02 & 0.01 & &   &   & 0.52 & 0.04 & 0.03 & &   &   & 1.52 & 0.30 & 0.25 &   &   & & 4.72 & 0.93 & 0.76  \\
        T6 & 0.17 & 0.03 & 0.02 & &   &   & 0.66 & 0.21 & 0.19 & &   &   & 1.84 & 0.47 & 0.41 &   &   & & 5.69 & 0.32 & 0.30  \\
        T7 & 0.84 & 0.30 & 0.27 & &   &   & 0.62 & 0.05 & 0.04 & &   &   & 3.07 & 1.35 & 1.10 &   &   & & 6.49 & 1.24 & 1.14  \\
        L1 & 0.43 & 0.02 & 0.02 & &   &   & 0.32 & 0.01 & 0.01 & &   &   & 0.87 & 0.09 & 0.09 &   &   & & 0.77 & 0.04 & 0.03  \\
        L2 & 0.77 & 0.14 & 0.12 & &   &   & 0.36 & 0.01 & 0.01 & &   &   & 1.18 & 0.05 & 0.05&   &   &  & 1.01 & 0.04 & 0.03  \\
        L3 & 1.15 & 0.04 & 0.03 & &   &   & 0.37 & 0.02 & 0.02 & &   &   & 1.26 & 0.12 & 0.11 &   &   & & 1.07 & 0.08 & 0.07  \\
        L2-10 & 0.44 & 0.13 & 0.09 & &   &   & 0.36 & 0.04 & 0.03 & &   &   & 1.66 &  0.12 & 0.09 &   &   & & 1.28 & 0.09 & 0.06  \\
        T1-10 & 0.027 & 0.007 & 0.005 & &   &   & 0.39 & 0.06 & 0.05 & &   &   & 1.28 &  0.18 & 0.14 &   &   & & 4.63 & 0.95 & 0.73  \\
        T1-10c & 0.037 & 0.006 & 0.006 & &   &   & 0.39 & 0.07 & 0.05 &   &   & & 1.08 & 0.05 & 0.04 &   &   & & 3.60 & 0.27 & 0.25 \\
        T1-r & 0.015 & 0.002 & 0.002 & 19.5 & 2.0 & 1.8 & 0.65 & 0.16 & 0.14 & 27.4 & 1.6 & 1.4 & 49.2 & 12.7 & 10.5 & 3.9 & 1.7 & 1.4 & 83.9 & 6.4 & 5.3 \\
        T1-10r & 0.016 & 0.009 & 0.007 & 21.3 & 2.3 & 1.9 & 0.55 & 0.16 & 0.13 & 29.5 & 2.7 & 2.4 & 69.0 & 24.5 & 21.9 & 4.8 & 2.8 & 2.4 & 64.1 & 13.8 & 11.4 \\
        L2-sim & 0.600 &   &   &  &   &   & 0.35 &   &   &   &   &   & &   &   & 1.4 &   & &&&   \\
        T1-sim & 0.050 &   &   &  &   &   & 0.40 &   &   &   &   &   & &   &   & 2.4 &   & &&&   \\
        T1r-sim & 0.015 &   &   &  &   &   & 5.00 &   &   &   &   &   & &   &   & 10.0 &   & &&&   \\
        \hline    

        \end{tabular}

        \begin{tabular}{|cccccccccccccc|}
         \hline
           & $s$ [mm] & $\alpha^-$ &       & $\alpha^0$  &       & $\alpha^+$    &      & $\beta^-$     &        & $\beta^0$   &       & $\beta^+$     &  \\           
        ID & & avg.       & std.  & avg.      & std.      & avg.          & std.  & avg.         & std.  & avg.      & std.  & avg.          & std.\\
        T1 & 20 &          &       & 2161      & 553       &               &       & 6536         & 582   &           &       & 19749         & 512\\ 
        T2 & 20 &          &       & 1996      & 1966      &               &       & 3830         & 1945  &           &       & 14588         & 1922  \\ 
        T3 & 20 &          &       & 492       & 122       &               &       & 1379         & 135   &           &       & 4551          & 114  \\ 
        T4 & 20 &          &       & 563       & 185       &               &       & 1794         & 169   &           &       & 4595          & 180  \\ 
        T5 & 20 &          &       & 320       & 35        &               &       & 936          & 59    &           &       & 2912          & 59   \\ 
        T6 & 20 &          &       & 381       & 116       &               &       & 1062         & 64    &           &       & 3290          & 46  \\ 
        T7 & 20 &          &       & 75        & 25        &               &       & 368          & 36    &           &       & 777           & 28  \\ 
        L1 & 20 &          &       & 75        & 4         &               &       & 202          & 8     &           &       & 179           & 4  \\ 
        L2 & 20 &          &       & 47        & 7         &               &       & 154          & 7     &           &       & 131           & 7  \\ 
        L3 & 20 &          &       & 32        & 2         &               &       & 109          & 3     &           &       & 93            & 2  \\ 
        L2-10 & 10 &       &       & 20        & 4         &               &       & 94           & 4     &           &       & 73            &  4 \\ 
        T1-10 & 10 &       &       & 363       & 83        &               &       & 1182         & 82    &           &       & 4274          & 92   \\ 
        T1-10c & 10 &       &       & 264       & 54        &               &       & 735          & 41    &           &       & 2454          & 44  \\ 
        T1-r & 20 &126035   & 581   & 4177      & 1017      & 177331        & 475   & 318437       & 989   & 25440     & 1403  & 543714        & 502 \\ 
        T1-10r & 10 &34094  & 406   & 881       & 453       & 47290         & 405   & 110535       & 487   & 7750      & 587   & 102794        & 428 \\ 
        T1-10cr & 10 &44109 & 334   & 2619      & 698       & 58775         & 329   & 146168       & 271   & 20272     & 240   & 174718        & 239\\ 
        L2-sim & 10 &      &       & 15        &           &               &       &              &       & 59        &       &               &  \\ 
        T1-sim & 10  &     &       & 200       &           &               &       &              &       & 1200      &       &               &  \\ 
        T1r-sim & 10 &     &       & 8333      &           &               &       &              &       & 16667     &       &               &  \\
        \hline
        \end{tabular}        
        
        
        \caption{
        \textbf{Overview of measured stiffness values}. Bending, shearing, compression, and extension stiffnesses are presented for all hinges (living and textile, embedded in rubber and plastic) presented in this work. For hinges in rubber matrices,  $k_b$, $k_s^0$ and $k_l^{\pm}$ are reported; for hinges embedded in plastic,  $k_b$, $k_s^0$, $k_s^{\pm}$, $k_l^0$, and $k_l^{\pm}$ are shown. Values used for simulations of arrays of blocks with typical side length $s=10$mm are shown in rows labelled L2-sim, T1-sim, and T1r-sim. Columns distinguish average, range, and standard deviation of all measurements. Shear and stretch parameters $\alpha, \beta$---quantifying mechanism-like behaviour---and their corresponding standard deviations are reported (Eqs.~\ref{eq:alphabeta}-~\ref{eq:alphabeta_error}.)
                    }
        \label{table:SI_testing_hinges}
    \end{table}

   \subsection{Arrays of multiple hinged units}
\begin{figure}[h]
        \begin{center}        
        \includegraphics[width=0.5\linewidth]{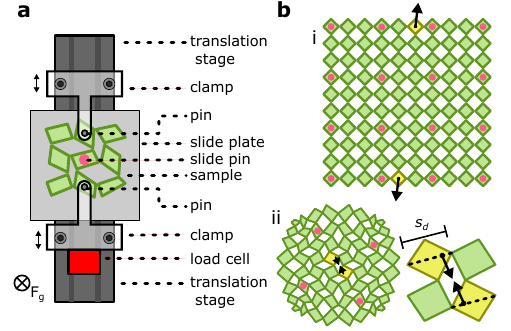}
            \caption{
            \textbf{Mechanical testing on larger arrays.}
            \textbf{a)} Schematic illustration of the two-point testing setup used for arrays of connected quadrilaterals. A sample is pinned horizontally to two opposing clamps, and rests on a slide plate. Rounded slide pins are attached to the sample to reduce friction with the plate. The clamps move along a motorized translation stage; during motion, axial forces are recorded with a load cell.
            \textbf{b)} Sample geometries (green squares), sliding pin locations (pink circles), and actuation points and directions (black arrows) are shown for the architectures used in this work. Schematics correspond to Fig.~3 (i) and Fig.~5(ii). Note that blocks in configuration $(ii)$ are subject to loading at distance $0.2 s_d$ from their corner vertices where $s_d$ is the blocks' diagonal length, while blocks in configuration $(i)$ are actuated at their geometric centers.
            }
            \label{fig:SI_testing_arrays}
            \vspace{-15pt}
        \end{center}
    \end{figure}

     Arrays of hinged units were tested using the setup shown schematically in Fig.~\ref{fig:SI_testing_arrays}a. Two translation stages (Thorlabs LTS300 and LTS150), mounted horizontally, are outfitted with sample clamps. One clamp is connected to a $50$-lb load cell (Futek LSB200). Samples are pinned between clamps using one small dowel pin, resulting in point forcing. Pins with rounded heads are inserted into the sample and allow it to rest on a slide plate lubricated with WD-40 to minimize friction forces and to prevent sagging of the sample under gravity. The translation stages and load cell are controlled with a custom Python library, allowing controlled displacements while forces are recorded. The speed of displacement is set at $0.5$ mm/s throughout.\\
    
     Fig.~\ref{fig:SI_testing_arrays}c shows the two array architectures tested in this work, along with their forcing directions (yellow squares, black arrows) and slide pin locations (pink dots). Fig.~\ref{fig:SI_testing_arrays}c(i) illustrates a 10-by-10 array of squares (side length $s=10$ mm) supported by 16 slide pins, actuated at the geometric centers of two edge squares. Fig.~\ref{fig:SI_testing_arrays}c(ii) shows a 10-by-10 array of irregular quadrilaterals supported by four slide pins, actuated at two central blocks. In each block, the actuation point was chosen to lie on a diagonal line (length $s_d$), at a distance of $0.2 s_d$ from the corner vertex.

   \subsection{Fracture testing on arrays of multiple hinged units}

\begin{figure}[h]
        \begin{center}        
        \includegraphics[width=0.5\linewidth]{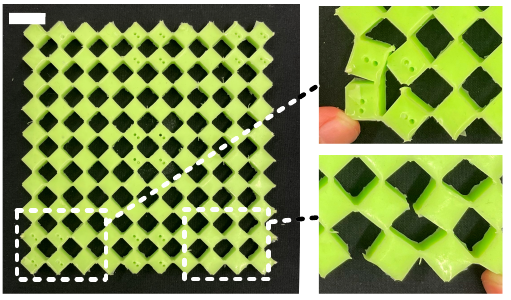}
            \caption{
            \textbf{Failure during demoulding in arrays of living hinges.} Demoulding of a 10-by-10 array of squares connected by living hinges of type L2 (thickness $1.5 \pm 0.2$ mm) resulted in fracture of 4 out of 100 hinges (zoom-ins, right).
            Scale bar: $20$ mm.
            }
            \label{fig:SI_array_break}
            \vspace{-15pt}
        \end{center}
        \end{figure}

   As shown in Fig.~\ref{fig:SI_array_break}, demoulding of 10-by-10 arrays of rubber squares connected by living hinges $L2$ resulted in fracture of several hinges. No such breakage was observed for any arrays of regular or irregular rubber or plastic quadrilaterals connected by textile hinges.

\begin{figure}[h]
        \begin{center}        
        \includegraphics[width=0.95\columnwidth]{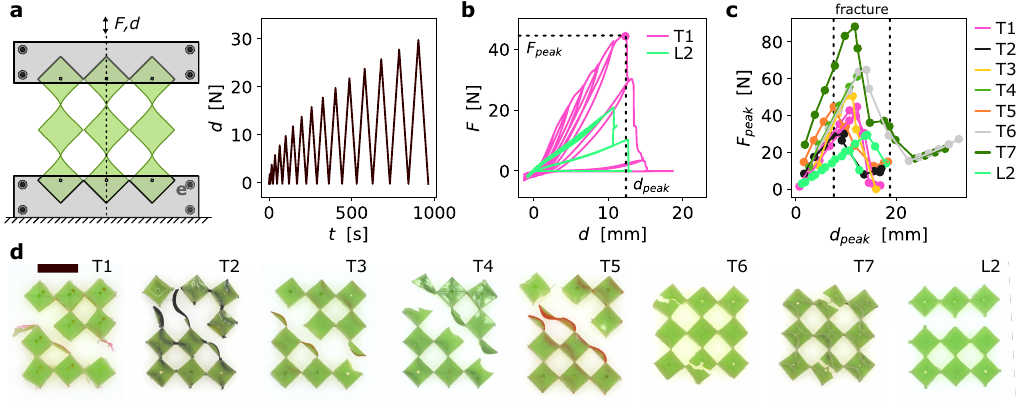}
            \caption{
            \textbf{Tensile fracture testing of rubber arrays with varying hinge types.}
            \textbf{a)} Left: schematic testing setup. Three-by-three samples of $s=10$-mm wide, $h=6.3\pm0.5$-mm thick squares are pinned to and bolted between plates with square cutouts. Clamps are secured in a testing machine that applies extension $d$ to the sample while monitoring load $F$. Right: the applied displacement protocol cycles the sample to increasing extensions over time $t$ at speed $0.5$mm/s.
            \textbf{b)} Typical force-displacement results for samples with textile hinges $T1$ and living hinges $L2$ (legend). After several cycles, sequential failure after reaching peak force $F_{peak}$ at displacement $d_{peak}$ (signified by abrupt force drops) are observed. 
            \textbf{c)} Peak force and peak displacement per cycle for all hinge types (legend) are plotted. Displacement at break (delineated by dashed lines) is similar across samples. Compared to living hinge $L2$, peak force at failure for textile hinges are similar or larger up to a factor four.  
            \textbf{d)} Snapshots of samples post-failure. Textile hinges T1 through T5 show dehesion between the textile hinge and rubber matrix along diagonals. Hinges T6 and T7 (with high cotton content) result in fracture of the rubber squares rather than dehesion. Living hinges L2 consistently break along adjacent horizontal hinges. Scale bar: 20mm.
            }
            \label{fig:SI_fracturing_arrays}
            \vspace{-15pt}
        \end{center}
        \end{figure}

   Here, we explore the comparative robustness of textile and living hinges by performing tensile fracture tests on 3-by-3 arrays of rubber squares connected by living hinges L2 and textile hinges T1 through T7. Fig.~\ref{fig:SI_fracturing_arrays}a shows the schematic testing setup: samples are pinned to plates with partial square cutouts and clamped with nuts and bolts.  Clamps are connected to an Instron 5969 universal testing machine, outfitted with a 2530-series $50$-N load cell, which applies displacement $d$ to the sample at $0.5$mm/s while monitoring response force $F$. Cyclic increasing displacement from $2$ to $30$ mm was applied over $15$ cycles. Fig.~\ref{fig:SI_fracturing_arrays}b shows representative force-displacement measurements for samples with hinge $L2$ and $T1$. Fracture events, corresponding to rapid drops in force, are observed over several cycles after $d\approx10$mm. To further explore fracture behaviour, peak force $F_{peak}$ and corresponding displacement $d_{peak}$ for each cycle were measured and plotted in Fig.~\ref{fig:SI_fabrication_arrays}c for all hinge types. Applied displacements at fracture are similar across all samples. However, peak forces at fracture for living hinges L2 are similar to or smaller than those for textile hinges, indicating increased robustness of textile hinges under high loads. 
   
   Finally, representative snapshots of all samples after testing are shown in Fig.~\ref{fig:SI_fracturing_arrays}d. Samples with textile hinges fracture diagonally along the fabric's weaving direction. They either exhibit dehesion between the textile material and rubber matrix (T1-T5) or fracturing of the rubber matrix near the textile inclusion (T6-T7). Note that the former behaviour occurs for synthetic textiles with smooth fibers, while the latter occurs for cotton-rich textiles with rough fibers, suggesting that rough fibers increase adhesion between the hinge and matrix material. Finally, living hinges fracture sequentially at horizontally adjacent hinges, where stress concentrations are highest. The fracture mode can therefore be influenced by the weaving pattern of the textile hinges.

\clearpage
\newpage

\section{Simulations}

\subsection{Mathematical model}

\begin{figure}[htb]
    \begin{center}
    \includegraphics[width=0.95\columnwidth]{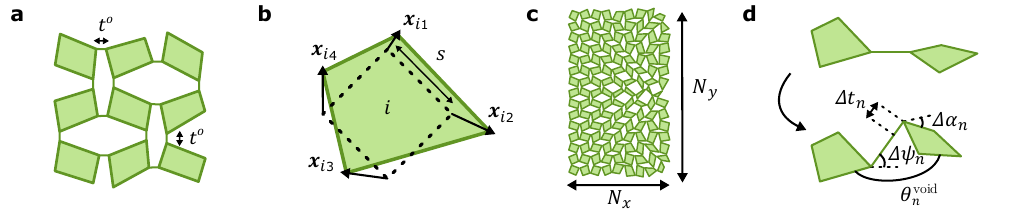}
        \caption{
        \textbf{Numerical model.}
        \textbf{a.} The model consists of two-dimensional rigid quadrilaterals connected by straight flexible ligaments of rest length $t^0$.
        \textbf{b.} The shape of the $i$-th quadrilateral is set by shifts $x_{i1}$ to $x_{i4}$ (arrows) of its four corner nodes away from a reference square with side length $s$ (dashed lines). 
        \textbf{c.} Example of a square array of $N_x$-by-$N_y$ quadrilaterals.
        \textbf{d.} Deformations of the $n$-th ligament are decomposed into bending, stretching, and shearing via deformations $\Delta \alpha_n$, $\Delta \psi_n$ and $\Delta t_n$ away from its rest configuration. Contact between blocks connected by the ligament is identified by tracking the void angle $\theta_n^{\mathrm{void}}$.
        }
        \label{fig:SI_modeling}
    \end{center}
\end{figure}

Here, we describe the numerical model used to predict the behaviour of arrays of quadrilaterals connected by thin hinges. As recently shown~\cite{deng_elastic_2017,coulais_characteristic_2018}, such systems can be accurately modeled as two-dimensional quadrilateral rigid blocks of side length $s$ connected at their vertices through thin ligaments of fixed length $t^0$ to form square arrays (Fig.~\ref{fig:SI_modeling}a). 
 The shape of each block  is fully described by the positions of its four corner nodes. As shown in Fig.~\ref{fig:SI_modeling}b,
 such positions for the $i$-th
block are defined by the vector $\vec{x}_i = (\vec{x}_{i1}, \vec{x}_{i2}, \vec{x}_{i3}, \vec{x}_{i4})$,  where $\vec{x}_{ip}$ denotes the displacement along the x and y direction of the
$p$-th  vertex of the $i$-th block from the corresponding vertex of a square with edge of $s = 10$ mm (highlighted with a black dashed line  in Fig.~\ref{fig:SI_modeling}b) and $t^0=0.05 s$. In Fig.~~\ref{fig:SI_modeling}c, we show an example of an $N_x$-by-$N_y$ array of irregular quadrilaterals, obtained by shifting the corner nodes of an array of regular squares by overall shift vector $\vec{x} = (\vec{x}_1...\vec{x}_i...\vec{x}_{N_xN_y})$.
Note that, to ensure that all ligaments have the same rest length $t^0$ across the non-periodic array, additional constraints are imposed on shift variables $\vec{x}_i$. Specifically, for  block $i$ connected to block $j$ via a ligament at nodes $k$ and $l$ we impose $\vec{x}_{ik}=\vec{x}_{jl}$.\\

Deformations of the array away from its resting configuration cost energy (Fig.~\ref{fig:SI_modeling}d). We  take into account ligament deformations and block contacts as follows. 
First, we model contact between neighbouring rigid blocks by introducing a contact energy, $\mathcal{E}_\text{con}$. Such contact energy is assumed to be  only active when the angle between connected blocks $\theta_n^\text{void}$ is below a cutoff angle $\theta_\text{cutoff}=\SI{5}{\degree}$:
\begin{equation}
\label{eq:energy_contact}
    \mathcal{E}_\text{con} = \sum_{n} \frac{k_\text{c}}{2} \left(\theta_n^\text{void}-\theta_\text{cutoff}\right)^2 \left(1-\tau_n^2\right)^{-1} \mathcal{H}(\theta_\text{cutoff}-\theta_n^\text{void})~,
\end{equation}
where $\mathcal{H}$ denotes the Heaviside function, and $\tau_n=(\theta_n^\text{void}-\theta_\text{cutoff})/(\theta_\text{cutoff}-\theta_\text{min})$. Furthermore, $k_\text{c}=100\,k_b$ controls the initial contact stiffness, and $\theta_\text{min}=\SI{0}{\degree}$ is the position of the vertical asymptote acting as an infinite energy barrier.
Second, the deformation of each ligament $n$ is decomposed into bending, longitudinal stretching, and shearing, quantified by $\Delta\alpha_n$, $\Delta t_n$, and $\Delta \psi_n$, respectively. It follows that the energy contribution of the ligaments reads as~\cite{coulais_characteristic_2018}
\begin{equation}
\label{eq:energy_ligaments}
    \mathcal{E}_\text{lig} = \frac{1}{2}\sum_{n} k_b\Delta\alpha_n^2 + k_l \Delta t_n^2 + k_s \left(t^0\Delta\psi_n\right)^2~,
\end{equation}
where  $k_b$, $k_l$, and $k_s$ are stiffness values. In this work, $k_b$, $k_l$, and $k_s$ are chosen to match the experimental measurements (see Table~\ref{table:SI_testing_hinges}). For samples of thickness $h=6.3\pm0.5$mm, we set $(k_b, k_s, k_l)=(0.6\,\mathrm{mNm},\, 0.35\,\mathrm{N/mm},\, 1.4\, \mathrm{N/mm})$ for hinges L2;  $(0.05\,\mathrm{mNm},\, 0.4\,\mathrm{N/mm},\, 2.4\, \mathrm{N/mm})$ for hinges T1; and $(0.015\,\mathrm{mNm},\, 5\,\mathrm{N/mm},\, 10\, \mathrm{N/mm})$ for hinges T1-r. For samples of thickness $h=4\pm0.5$mm, all stiffnesses (which scale linearly with $h$) are reduced by a factor $4/6.3$ for hinges L2; by contrast, only shearing and stretching stiffnesses for hinges T1 and T1-r are reduced by this factor. \\

To simulate the response of the metamaterial upon quasi-static loading, we adopt the method described in~\cite{bordiga_2024}.
The strategy consists of a dynamic three-step approach.
First, we construct the potential energy function of the system 
\begin{equation}
\mathcal{E}_\text{tot}=\mathcal{E}_\text{lig}+\mathcal{E}_\text{con}~.
\end{equation}
Second, we derive the equations of motion by taking the gradient of the potential energy with respect to the degrees of freedom:
    \begin{equation}\label{eq:equations_of_motion}
    \bm{M}\ddot{\vec{u}} = -\frac{\partial \mathcal{E}_\text{tot}(\vec{u})}{\partial\vec{u}} -\bm{C}\dot{\vec{u}} ~,
    \end{equation}
where $\bm{M}=\text{diag}(m_1,m_1,I_1,...,m_{N_x N_y},m_{N_x N_y},I_{N_x N_y})$ is the inertia matrix of the system, and $\bm{C}\dot{\vec{u}}$ is a linear viscous damping term. Moreover, $\vec{u}=(u^x_1,u^y_1,\theta_1,...,u^x_{N_x N_y},u^y_{N_x N_y},\theta_{N_x N_y})$ is a vector with all degrees of freedom of the metamaterial.
Lastly, we integrate the equations of motion  numerically, using a Dormand--Prince explicit solver with adaptive stepsize~\cite{dormand_1980} under slowly evolving boundary conditions. As we are seeking the quasi-static response of the system, we can artificially tune mass and damping to speed up the numerical integration of Eq.~\eqref{eq:equations_of_motion}.
In all simulations, the mass of the units is computed using a density $\rho=\SI{1.2e3}{kg/m^2}$, while the damping coefficients are assumed as $c_x=c_y=0.1 \sqrt{k_l \rho s^2\sqrt{2}}$, $c_\theta=0.1\sqrt{k_l\rho s^4\sqrt{2}}$ where $s=\SI{10}{mm}$ is the edge length of the reference square unit.\\

Note that these steps are all implemented in a custom Python library, making use of the automatic differentiation package JAX~\cite{bradbury_2018}.
The package allows us to streamline the definition of the energy functional $\mathcal{E}_\text{tot}(\vec{u})$, the computation of its gradient $\partial \mathcal{E}_\text{tot}(\vec{u})/\partial\vec{u}$, as well as the numerical integration of the equation of motions.

\subsection{Inverse design}

\begin{figure*}[ht]
	\centering
	\includegraphics[width=\linewidth]{{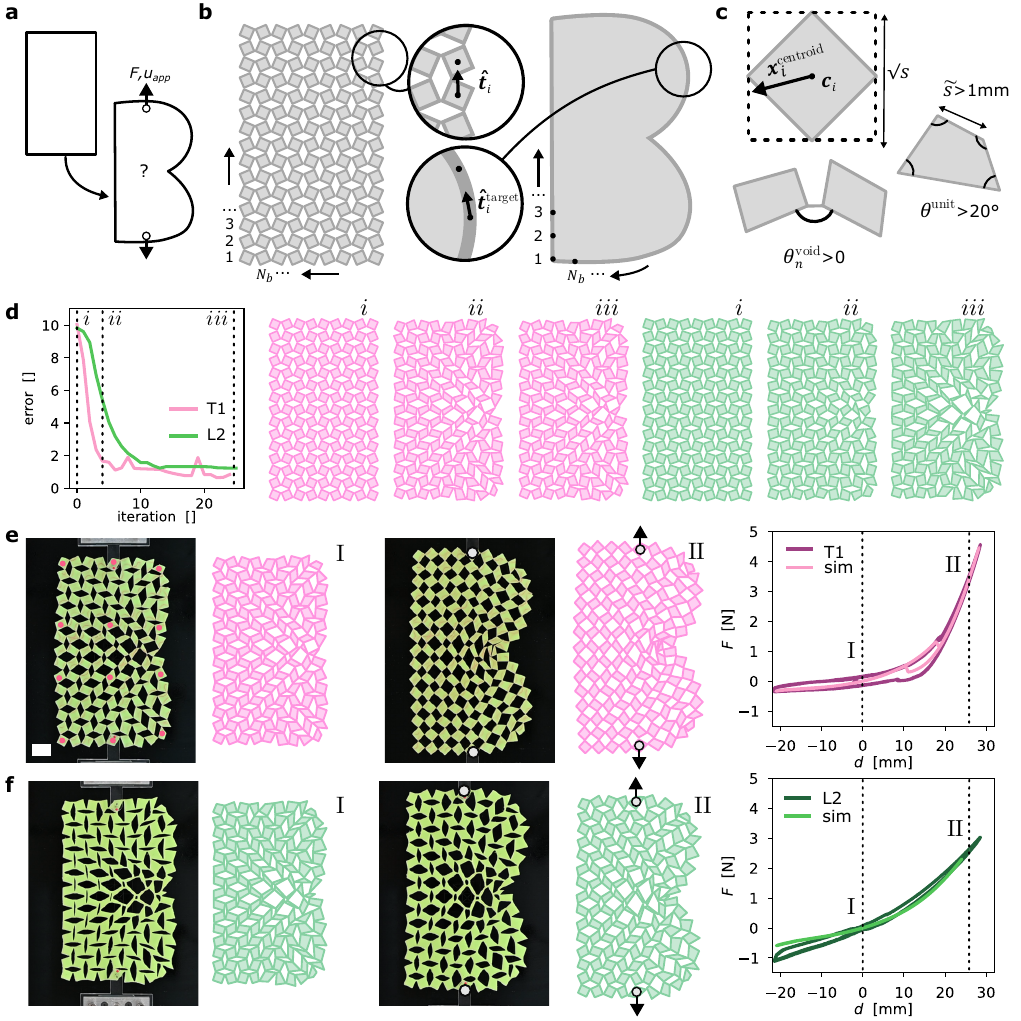}}
	\caption{
		\textbf{Inverse design of shape-morphing metamaterials with living or textile hinges.}
		\textbf{a)} Design goal: an initially rectangular metamaterial morphs to a target boundary shape under applied displacement $u_{app}$. 
          \textbf{b)} Defining a boundary shape. The shape of the initial array is found by calculating unit vectors $\uvec{t}_i$ between the centroids of $N_b$ neighbouring boundary blocks. The target shape is determined by equally spacing $N_b$ target centroids around the target boundary and calculating unit vectors $\uvec{t}_i^{\mathrm{target}}$ between consecutive pairs.
        \textbf{c)} Constraints during optimization. Left: the centroid $\vec{c}_i$ of the $i$-th block shifts by $\vec{x}_i^{\mathrm{centroid}}$. The maximal shift is limited by $-\frac{1}{2}\sqrt{s}<x^{\mathrm{centroid}}_{i~x,y} < \frac{1}{2}\sqrt{s}$ (dashed lines). Top right: neighbouring blocks may not overlap, which is guaranteed by $\theta_n^{\mathrm{void}} >0\si{\degree}$. Bottom: blocks may not have acute internal angles $\theta^{\mathrm{unit}}$ below $20\si{\degree}$.
		\textbf{d)} Left: error evolution during optimization, using textile hinges T1 and living hinges L2 in a rubber matrix of 9 by 15 quadrilaterals. Right: snapshots of evolving metamaterial geometry ($i-iii$).
		\textbf{e)} Experimental and simulated testing of optimized design for textile hinges T1. Left: sample at rest ($I$). Pink dots indicate sliding pin locations (see Fig.~\ref{fig:SI_testing_arrays}). Middle: sample under extension ($II$). Right: force-displacement curve. Scale bar: 20mm.
        \textbf{f)} Experimental and simulated results of optimized design with living hinges L2. Results show excellent agreement between model and experiment.
	}
	\label{fig:SI_Bmorpher}
\end{figure*}

The numerical model above enables us to efficiently determine the  response of arrays of arbitrary hinged units. While it has been recently shown that the shape of the quadrilaterals has a profound effect on the nonlinear mechanical response of the resulting metamaterials~\cite{Deng2022}, the connection between the geometry of the units and the change in shape of the metamaterial upon application of loading is not trivial. As such,  identification of metamaterials capable of shape-morphing into target shapes upon application of a given load requires an efficient inverse design strategy. We now describe the inverse design framework used to discover metamaterial geometries capable of achieving target deformations upon loading as presented in previous work~\cite{bordiga_2024}. The framework makes use of the fact that, given any prescribed quasi-static loading, the numerical integration of Eq.~\eqref{eq:equations_of_motion} rapidly provides the deformation of a given metamaterial design described by the shift vector $\vec{x}$.\\

As an illustrative example, illustrated in Fig.~\ref{fig:SI_Bmorpher}a, we search for a metamaterial comprising a $9\times15$ array of hinged quadrilateral units connected at their vertices by textile
hinges $T1$ or living hinges $L2$ that morphs from a rectangle into a B letter upon application of a total vertical elongation $u_\text{app}=2s\sqrt{2}\approx\SI{28}{mm}$ to two pinned units at middle of the top and bottom edges.

We first choose and describe the target "B"-shape as follows. We implement a custom Python code (using the open source library OpenCV~\cite{opencv_library}) that takes an arbitrary binary image of a desired shape, extracts its boundary, downsamples the boundary pixels to the number of desired boundary blocks, and computes the  tangent unit vector to the boundary of the target shape, $\uvec{t}_i^\text{target}$, at each of these points. This strategy, illustrated in Fig.~\ref{fig:SI_Bmorpher}b, eliminates the need for explicit (and potentially overly complex) parametrization of the target shape.\\

Subsequently, to identify metamaterial geometries capable of matching the target shape upon loading, we seek to minimize the following objective function
\begin{equation}
    \label{eq:cost_function}
    J(\vec{x}) = \sum_{i\in\text{boundary}} \left\lVert \uvec{t}_i(\vec{x}) - \uvec{t}_i^\text{target} \right\rVert^2 ~,
\end{equation}
where $\uvec{t}_i(\vec{x})$ is the tangent unit vector to the boundary at the $i$-th block in the deformed configuration of the design described by the 540-dimensional vector $\vec{x}$.
Importantly, the objective function above is scale-invariant due to normalization of the tangent vectors. This allows the optimization algorithm to discover the best overall size of the deployed shape under the imposed loading, without having to specify a target size.\\

Finally, we note that the following geometric constraints are imposed on the design space $\vec{x}$, as shown in Fig.~\ref{fig:SI_Bmorpher}c. First, the centroid of the $k$-th block is subject to a box constraint
    \begin{equation*}
        \lvert x^\text{centroid}_k\rvert\leq\sqrt{s}/2\,, \quad \lvert y^\text{centroid}_k\rvert\leq\sqrt{s}/2
    \end{equation*}
in order to confine the design space, accelerating the optimization while also providing a rest configuration with a nearly rectangular boundary shape.
Second, the void angle between each pair of connected blocks is constrained by
    \begin{equation*}
        \theta^\text{void}_{n,0}(\vec{x}) > 0 \si{\degree}
    \end{equation*}    
in order to avoid unfeasible overlapping designs.
Finally, the solid angles and the length of the sides of each quadrilateral block are constrained by
    \begin{align*}
        \theta^\text{unit}(\vec{x}) &> \SI{20}{\degree} \\
        \tilde{s}(\vec{x}) &> \SI{1}{mm}
    \end{align*}    
to avoid acute corners and excessively small edges, which are challenging to fabricate.\\

To efficiently search through the large space of metamaterial geometries, a gradient-based optimization strategy is implemented to minimize the objective function of Eq.~\eqref{eq:cost_function}. In particular, gradients of the objective function and the constraints are computed by leveraging the automatic differentiation package JAX~\cite{bradbury_2018} and design updates are performed via the Method of Moving Asymptotes (MMA)~\cite{svanberg_1987} (provided by the NLopt library~\cite{johnson_2007}).
All the code developed for the simulations and optimizations is available at \href{https://github.com/bertoldi-collab/DifFlexMM}{github.com/bertoldi-collab/DifFlexMM}.

As shown in Fig.~\ref{fig:SI_Bmorpher}d, our optimization algorithm quickly evolves the initial design, and after about 20 iterations, it identifies designs for both textile hinges T1 and living hinges L2 that morph into the target shape upon application of the desired loading. The optimized designs are characterized by units with a high aspect ratio in the middle-to-right region of the domain, especially close to the sharp cusp of the B shape, while top-left and bottom-left regions have a higher material-to-void ratio (Fig.~\ref{fig:SI_Bmorpher}d-\textit{iii}). The optimized designs are fabricated and tested experimentally. Fig.~\ref{fig:SI_Bmorpher}e and Fig.~\ref{fig:SI_Bmorpher}f show the resulting findings for the metamaterial with textile hinges T1 and living hinges L2, respectively. We show the experimental and modelled stress-free (Fig.~\ref{fig:SI_Bmorpher}e-f, left) and deployed configurations (Fig.~\ref{fig:SI_Bmorpher}e-f, middle), demonstrating the programmed shape transformation from rectangle to the letter B. The modelled and experimental deployed geometries are in good agreement across the whole structure, including the cusp region on the right edge. The experimental and simulated force-displacement curves (Fig.~\ref{fig:SI_Bmorpher}e-f, right) reveal a good match. Note that our optimization for textile hinges T1 discovered a design with a small snap-through instability during deployment.
The reason for this peculiar feature can be understood by examining the shape transformation as the sample is loaded: the snap-through triggers a sudden horizontal contraction in the middle of the sample thus allowing for a better match of the sharp cusp point on the right edge of the target B shape (see Movie S1).

\newpage
\clearpage

\section{Supporting Movies}

\emph{Movie S1: } A textile-hinged tubular structure capable of morphing between a cylinder and a cone~\cite{choi2021compact} (Fig.3g-h). We fabricate the $10\times10$ design using rubber quadrilaterals and textile hinges $T1$. As intended, this structure smoothly transitions between the target cylindrical and compact conical configurations).

\emph{Movie S2: } A previously identified kinematic design that transitions from an open circle to a compact square~\cite{choi2019programming} (Fig.5). In the mechanism limit, the design is predicted to be stable in its open and closed configurations. The $10\times10$ design is realized using living hinges L2 in a rubber matrix; textile hinges T1 in a rubber matrix; and textile hinges T1 in a plastic matrix. Experiments confirm that the latter sample, which most closely approaches mechanism-like behaviour by using textile hinges and rigid blocks, is bistable. 

\emph{Movie S3: } Experimental and simulated response of a shape-morphing metamaterial under extension $d$  while loading force $F$ is monitored (Fig.S6e). The structure, made out of a 9-by-15 array of rubber quadrilaterals connected by textile hinges T1, morphs from a rectangular to a "B"-shaped configuration. A small snap-through event is observed in both experiment and simulation.

\bibliographystyle{MSP}
\bibliography{./references/designerhinge}
\end{document}